\documentclass[a4paper]{jpconf}
\usepackage{graphicx}
\usepackage{amsmath}
\usepackage{amssymb}
\newcommand{\nn}{\nonumber}
\newcommand{\half}{\frac{1}{2}}
\begin{document}
\title{A Review of Noncommutative Field Theories
\footnote{Talk presented at the XIV Mexican School on Particles and Fields, Morelia, November 9-11, 2010 }
}

\author{Victor O. Rivelles}

\address{Instituto de F\'{i}sica, Universidade de S\~ao Paulo, Rua do Mat\~ao Travessa R 187,  05508-090 S\~ao Paulo, Brazil}

\ead{rivelles@fma.if.usp.br}

\begin{abstract}
We present a brief review of selected topics in noncommutative field theories ranging from its revival in string theory, its influence on quantum field theories, its possible experimental signatures and ending with some applications in gravity and emergent gravity. 
\end{abstract}

\section{Introduction}

The uncertainty principle of quantum mechanics $\Delta x^i \Delta p^j > \delta^{ij} \hbar$ teach us that if we want to probe short distances we need high energies. Now general relativity teach us that if we keep increasing the energy to explore smaller regions a black hole will be formed as soon as we goes beyond the Schwarzschild radius. Since no information can scape from a black hole it seems that there is a lower bound for measuring lengths! This means that there must exist an uncertainty relations for the coordinates 
\begin{equation}
	\Delta x^\mu \Delta x^\nu > \theta^{\mu\nu}, \nn
\end{equation}
where $\theta^{\mu\nu}$ has dimensions of $(\mbox{lenght})^2$, so that coordinates no longer commute. This simple argument was presented some years ago \cite{Doplicher:1994tu} but the idea of noncommutativity is much older. In the late 30´s Heisenberg was struggling with the ultraviolet divergences of QED and thought that noncommuting coordinates could provide a cutoff to regulate them. He then wrote to Peierls, who told to Pauli, who told to Oppenheimer, who asked one of his students, Snyder, to think about it. Then in 1947 Snyder wrote a paper where he considered the following commutation relations \cite{Snyder1}
\begin{align}
	[ x^\mu, x^\nu ] &= i \ell^2 \hbar^{-1} ( x^\mu p^\nu - x^\nu p^\mu), \nn \\
	 \left[ x^\mu, p_\nu \right] &= i \hbar \delta^\mu_\nu + i \ell^2 \hbar^{-1} p^\mu p_\nu, \nn \\
	\left[ p_\mu, p_\nu \right] &= 0, 
\end{align}
generalizing the usual ones and with a new constant $\ell$. However the renormalization program for QED started to become successful so further studies in noncommutativity were discouraged.

Even so we can ask how we can do field theory with noncommutative coordinates satisfying a  simpler set of commutation relations 
\begin{align}
	\left[ \hat{q}_i, \hat{p}_j \right] &= i \hbar \delta_{ij},  \nn \\
	\left[ \hat{q}_i, \hat{q}_j \right] &= i \theta_{ij}, \nn \\
	\left[ \hat{p}_i, \hat{p}_j \right] &= 0, \nn
\end{align}
with $\theta_{ij}$ a constant antisymmetric matrix. 
We can make an association of fields $\phi(x)$ in the usual space with commutative coordinates $x^i$  to operator valued objects $\hat{\Phi}(\hat{q})$ as 
\begin{align}
	\hat{\Phi}(\hat{q}) = \int dp \,\, e^{i p \hat{q} } \tilde{\phi}(p), \nn
\end{align}
where $\tilde{\phi}(p)$ is the usual Fourier transform of $\phi(x)$, $\tilde{\phi}(p) = \int dx \,\, e^{-ipx} \phi(x)$. After using the Hausdorff-Campbell formula for the product of noncommutative fields we get 
\begin{align}
	\hat{\Phi}_1(\hat{q}) \hat{\Phi}_2(\hat{q}) = \int dp_1 \,\, dp_2 \,\,
e^{i(p_1 + p_2) \hat{q} - \frac{1}{2}  p_1^\mu p_2^\nu
  \theta_{\mu\nu}} \phi_1(p_1) \phi_2(p_2), \nn
\end{align}
so that we can associate to this product a deformed product of commutative fields in ordinary space
\begin{align}
	\hat{\Phi}_1(\hat{q}) \hat{\Phi}_2(\hat{q}) \leftrightarrow \left( \phi_1
  \star \phi_2 \right) (x),\nn
\end{align}
know as the star or Moyal product
\begin{align}
\left( \phi_1 \star \phi_2 \right) (x) \equiv \left[ e^{i\frac{1}{2}
    \theta^{\mu\nu} \frac{\partial}{\partial x^\mu}
    \frac{\partial}{\partial y^\nu} } \phi_1(x) \phi_2(y)
\right]_{y=x} = \phi_1(x) \phi_2(x) + i \frac{1}{2}
    \theta^{\mu\nu} \partial_\mu \phi_1 \partial_\nu \phi_2 + \dots. \nn
\end{align}
In this way we can work in the usual space with commutative coordinates and conventional fields but replacing the ordinary product of fields by the Moyal product. We can even consider the quantization of such field theories and one remarkable result is that the structure of ultraviolet divergences is not modified by the Moyal product \cite{Filk:1996dm}. This is quite surprising because when the Moyal product is expanded there is an infinite number of higher derivative terms which contribute to the interaction vertices in Feynman diagrams. We would expect that the ultraviolet structure would be spoiled but the Moyal structure works in such a way as to preserve it. At the end, Heisenberg's dream of using noncommutativity to tame the divergences did not come true. But, as we shall see, other effects do appear and in fact are deadly for most field theories. 

Besides the importance of Filk´s work in 1996 it did not call much attention since only a few people were interested in noncommutativity on those days. The revolution in noncommutativity had to wait a few years more for another important discovery, but this time in string theory. Dp-branes are extended objects in $p$ space dimensions in which open strings can end. When a set of $N$ D-branes are nearly coincident the string spectrum includes a $U(N)$ gauge theory in the world-volume of the D-branes. It also includes a matrix model of $N\times N$ matrices for each transverse dimension of the brane. We can consider the situation when the NS-NS antisymmetric field $B_{\mu\nu}$ is turned on in the presence of D3-branes. It was found that there is a low energy limit where the closed strings decouple and the effective theory living on the D3-brane is a noncommutative gauge theory \cite{Seiberg:1999vs}. The noncommutativity is induced by the Moyal product with $\theta_{\mu\nu}$ being related to the NS-NS field $B_{\mu\nu}$. Since these noncommutative theories arouse from a consistent truncation of string theory they must be consistent in some sense and this gave rise to an intense period of research in noncommutative field theories. 

Since then the proposal of noncommutativity has been applied to many areas of physics and mathematics and would be impossible to review all advancements happened along these years. Instead I will concentrate on topics which I have been involved, mostly concerned with quantum field theory and gravity. A couple of older reviews are also useful as an introduction to the subject \cite{Douglas:2001ba,Szabo:2001kg}.

\section{Mixing of Divergences}

An important feature of quantum noncommutative field theories is the mixing of ultraviolet and infrared divergences usually known as the UV/IR mixing \cite{Minwalla:1999px}. Consider the  noncommutative scalar field theory 
\begin{align}
	S = \int d^4 x \,\, \left( \half \partial_\mu \hat{\phi} \star \partial^\mu \hat{\phi}
- \frac{m^2}{2} \hat{\phi} \star \hat{\phi} - \frac{g^2}{4!}
\hat{\phi}\star\hat{\phi}\star\hat{\phi}\star\hat{\phi} \right), \nn
\end{align}
where the ordinary product among the fields was replaced by the Moyal product. It easy to show that 
$\int dx \,\, (f \star g)(x) = \int dx \,\, f(x) g(x)$ so that the quadratic terms in the action are not affected by the Moyal product. This means that propagators are not modified in noncommutative theories. But the interaction vertex is modified
\begin{align}
 &\int d^4x \,\, \hat{\phi}\star\hat{\phi}\star\hat{\phi}\star\hat{\phi} = 
 -\frac{1}{3} \int dk_1 dk_2 dk_3 dk_4 \,\, \delta
(k_1+k_2+k_3+k_4)
\times [ \cos(\half k_1 \wedge k_2) \cos (\half k_3 \wedge k_4) + \nn\\ 
&+\cos(\half k_1 \wedge k_3) \cos (\half k_2 \wedge k_4) +  \cos(\half k_1 \wedge k_4) \cos (\half k_2 \wedge k_3) ] \,\,
\hat{\phi}(k_1) \, \hat{\phi}(k_2) \,  \hat{\phi}(k_3) \, \hat{\phi}(k_4), \nn
\end{align}
where $k \wedge p = \theta_{\mu\nu} k^\mu p^\nu$. This interaction leads to the the usual UV divergence which requires a UV regulator $\Lambda$. But it also gives rise to contributions which are singular in the IR. Symbolically the effective regulator has the form 
\begin{align}
	\Lambda_{eff}= \frac{1}{\frac{1}{\Lambda^2} + (\theta p)^2}, \nn
\end{align}
mixing with the UV with the IR divergence. This mixing does not spoils renormalizability at one loop but it is fatal at higher orders \cite{Minwalla:1999px}. Several proposals were made to overcome this trouble but all of them require non trivial modifications of the original theory \cite{Langmann:2002cc,Grosse:2004yu,Rivasseau:2005bh,Disertori:2006nq,Langmann:2003if}. The simplest solution, however, is the inclusion of supersymmetry which turns the noncommutative chiral multiplet renormalizable to all orders in perturbation theory \cite{Girotti:2000gc}. Other troubles caused by the UV/IR mixing may also be healed by supersymmetry like spontaneous symmetry breaking \cite{Girotti:2002kr} but other problems like the renormalizability of supersymmetric gauge theories  \cite{Ferrari:2004ex,Ferrari:2003vs} are not.  


The noncommutative abelian gauge theory is described by the action
\begin{align}
	S_A &= -\frac{1}{4} \int d^4x \,\,\, \hat{F}^{\mu\nu} \star
\hat{F}_{\mu\nu},\nn
\end{align}
with the field strength and gauge transformation modified by the addition of the Moyal commutator $\left[ \hat{A}, \hat{B} \right]_\star = \hat{A} \star \hat{B} - \hat{B} \star \hat{A}$ as
\begin{align}
\hat{F}_{\mu\nu} &= \partial_\mu \hat{A}_\nu - \partial_\nu
\hat{A}_\mu - i [ \hat{A}_\mu, \hat{A}_\nu ]_\star, \nn\\
\delta \hat{A}_\mu &= \hat{D}_\mu \hat{\lambda} = \partial_\mu \hat{\lambda} - i [\hat{A}_\mu, \hat{\lambda}]_\star. \nn
\end{align}
The noncommutative gauge field $\hat{A}_\mu$ can be mapped to the conventional abelian gauge field $A_\mu$, with the conventional field strength and gauge transformation, through the Seiberg-Witten map, which to first order in $\theta$ is  \cite{Seiberg:1999vs}
\begin{align}
	\hat{A}_\mu = A_\mu - \frac{1}{2} \theta^{\alpha\beta}
A_\alpha ( \partial_\beta A_\mu + F_{\beta\mu} ).  \nn
\end{align}
The resulting abelian gauge theory is now an interacting theory with action
\begin{align} \label{1}
	S_A = -\frac{1}{4} \int d^4x \,\,\, \left[ F^{\mu\nu} F_{\mu\nu} + 2
\theta^{\mu\rho} {F_\rho}^\nu \left( {F_\mu}^\sigma F_{\sigma\nu} +
\frac{1}{4} \eta_{\mu\nu} F^{\alpha\beta} F_{\alpha\beta} \right)
\right]. \nn
\end{align}
It gives rise to a nonrenormalizable theory due to the new interactions generated by the noncommutativity and also presents the UV/IR mixing. Its properties have been studied by several groups. Dualities among gauge theories usually do not survive after the Seiberg-Witten map \cite{Harikumar:2005ry,Ajith:2008qu}.

\section{Signatures of noncommutativity}

The existence of a constant antisymmetric tensor $\theta_{\mu\nu}$ means that Lorentz symmetry is broken. However, it does not mean that Lorentz symmetry is not valid at a fundamental level. We can think of $\theta_{\mu\nu}$ as a field (as in string theory) which has a constant vacuum expectation value breaking Lorentz symmetry spontaneously. So one of the manifestations of noncommutativity is through small deviations from Lorentz symmetry. The noncommutative Maxwell equations for the conventional abelian field  show that the photon has a modified dispersion relation \cite{Guralnik:2001ax} 
\begin{align}
	k^2 - 2 \theta^{\mu\alpha} {F_\alpha}^\nu k_\mu k_\nu = 0. \nn
\end{align}
This is similar to what happens to photons in an anisotropic medium with no Faraday rotation nor birefringence \cite{Mariz:2006kp}. Many other studies of modifies dispersion relations have also been performed \cite{Agostini:2003vg,Helling:2007zv,Bietenholz:2010dy,Bietenholz:2006cz}. 

There are several limits imposed on noncommutativity coming from many experiments. Observations of blazars give $|\theta| < (10^3 TeV)^{-2}$ \cite{AmelinoCamelia:2002dx}, atomic transitions in noncommutative quantum mechanics and Lamb shift in noncommutative QED give $|\theta| < (10 TeV)^{-2}$ \cite{Chaichian:2000si}, spin-statistics violations in noncommutative QED from Gran Sasso and Super-Kamiokande give $|\theta| < ( 10^{5} TeV )^{-2}$ \cite{Balachandran:2005eb}, noncommutative extensions of the standard model give $|\theta| < (10 TeV)^{-2}$ \cite{Carroll:2001ws} and noncommutative symplectic structure in classical mechanics and perihelium of Mercury give $|\theta|< (10^{13} TeV)^{-2}$ \cite{Romero:2003tu}. 

\section{Emergent Gravity}

There is a surprising connection between translations and gauge transformations. Consider a rigid translation for a noncommutative scalar field
\begin{align}
\delta_T \hat{\phi} = \xi^\mu \partial_\mu \hat{\phi}, \nn
\end{align}
and compare it with a noncommutative gauge transformation
\begin{align}
\delta \hat{\phi} = -i [ \hat{\phi}, \hat{\lambda} ]_\star 
= -i ( \hat{\phi} \star \hat{\lambda} - \hat{\lambda} \star
\hat{\phi} ). \nn
\end{align}
We can write the derivative of the field in terms of the Moyal commutator as $\partial_\mu \hat{\phi} = - i \theta^{-1}_{\mu\nu} [ x^\nu, \hat{\phi}]_\star$ assuming that $\theta$ is invertible, so that a translation can be written as a gauge transformation 
\begin{align} 
\delta_T \hat{\phi} = \delta \hat{\phi},\nn
\end{align}
with gauge parameter $\hat{\lambda} = - \theta^{-1}_{\mu\nu} \xi ^\mu x^\nu$. Therefore, translations in the noncommutative directions are equivalent to gauge transformations \cite{Gross:2000ph}. Of course, this has a flavor of general relativity! 

Let us consider this proposal in more detail. The action for a noncommutative scalar field coupled to a noncommutative abelian gauge field is given by 
\begin{align}
	S_\varphi = \frac{1}{2} \int d^4x \,\,\, \hat{D}^\mu \hat{\varphi} \star
\hat{D}_\mu \hat{\varphi}. \nn
\end{align}
We can now use the Seiberg-Witten map to work with the conventional scalar field $\varphi$ and the conventional abelian gauge field $A_\mu$ to get 
\begin{align}
	S_\varphi = \frac{1}{2} \int d^4x \, [ \partial^\mu \varphi \partial_\mu
\varphi + 2 \theta^{\mu\alpha} {F_\alpha}^\nu ( - \partial_\mu \varphi
\partial_\nu \varphi + \frac{1}{4} \eta_{\mu\nu} \partial^\rho \varphi
\partial_\rho \varphi ) ]. \nn
\end{align}
Notice that the tensor inside the parenthesis is traceless. Now consider the coupling of the conventional scalar field to linearized gravity 
\begin{align}
	S_{g,\varphi} = \frac{1}{2} \int d^4x \, \left( \partial^\mu \varphi
\partial_\mu \varphi - h^{\mu\nu}  \partial_\mu \varphi \partial_\nu
\varphi + h \partial^\rho \varphi \partial_\rho \varphi \right), \nn
\end{align}
where $h_{\mu\nu}$ is the traceless linearized metric and $h$ its trace. Comparing both actions we can identify the linearized gravitational field as
\begin{align}
	h^{\mu\nu} &=  \theta^{\mu\alpha} {F_\alpha}^\nu + \theta^{\nu \alpha}
{F_\alpha}^\mu + \frac{1}{2} \eta^{\mu\nu} \theta^{\alpha\beta}
F_{\alpha\beta},\nn \\
h &= 0. \nn
\end{align}
Therefore, the effect of noncommutativity on the scalar field is similar to a field dependent gravitational field. The coupling of matter to the abelian gauge field in noncommutative theories has effects that mimic those of gravity. In this way gravity is an emergent phenomena in noncommutative theories \cite{Rivelles:2002ez}. Remarkably charged fields feel a gravitational field that is half of that felt by the uncharged field so that the gravity coupling is now charge dependent. We can also compute the geometry induced by noncommutativity: it is that of a plane wave, more precisely a pp-wave \cite{Rivelles:2002ez}. These linearized results can be extended to the full theory \cite{Banerjee:2004rs}. As remarked in the introduction matrix models also appear when we consider D-branes in string theory. Again the emergence of gravity does arise in noncommutative  matrix models. The matrix-model for a noncommutative $U(N)$ gauge theory actually describes $SU(N)$ gauge theory coupled to gravity \cite{Steinacker:2007dq}.

\section{Noncommutative Gravity}

There are many attempts to formulate a noncommutative gravity theory leading to several extensions of general relativity. I will not try review all these proposals and will concentrate only in my attempt for such a formulation \cite{Harikumar:2006xf}. We tried to implement noncommutativity in curved spacetime in a manner which is as simple as possible. In flat spacetime $\theta_{\mu\nu}$ is a constant matrix. How could we consider it in curved spacetime? Before answering this question let us see what we know in flat spacetime. To this end let us assume that $\theta_{\mu\nu}$ is a tensor and perform an infinitesimal general coordinate transformation 
\begin{align}
 \delta \theta^{\mu\nu} = \xi^\lambda \partial_\lambda \theta^{\mu\nu} -
\partial_\lambda \xi^{\mu} \theta^{\lambda\nu} + \partial_\lambda
\xi^{\nu} \theta^{\lambda\mu}. \nn
\end{align}
Let us also assume that $ \delta \theta^{\mu\nu} = 0$. For a rigid translation $\xi^\mu$ is constant and we get that   $ \partial_\lambda \theta^{\mu\nu} = 0$, so that $ \theta^{\mu\nu}$ is  constant. Let us now consider a rigid Lorentz transformation $\xi^\mu = {\Lambda^\mu}_\nu x^ \nu$. This time we get ${\Lambda^\mu}_\lambda \theta^{\lambda\nu} - {\Lambda^\nu}_\lambda
\theta^{\lambda\mu} = 0$. Let us choose just one non-vanishing component of $\theta_{\mu\nu}$, say $\theta^{12} \not= 0$. Then Lorentz boosts in the $3$-direction and rotations in the $1-2$ plane are
still preserved as well as translations in any direction. We might think that these are all the solutions of $ \delta \theta^{\mu\nu} = 0$ but there is still one more. It is 
\begin{align}
\xi^\mu = \theta^{\mu\nu} \partial_\nu \xi, \nn
\end{align}
where $\xi$ is a scalar function. Notice that $\partial_\mu \xi^\mu= 0$.  This means that this transformation forms a symplectic subgroup of volume preserving diffeomorphisms  which also preserves  $\theta^{\mu\nu}$. This symmetry is not very useful in flat spacetime but will be very important in curved spacetime.

Let us now move to curved spacetime where the general coordinate transformation is 
\begin{align}
\delta \theta^{\mu\nu} = \xi^\lambda D_\lambda \theta^{\mu\nu} -
D_\lambda \xi^\mu \theta^{\lambda\nu} + D_\lambda \xi^\nu
\theta^{\lambda\mu}.
\end{align}
Assume again that $ \delta \theta^{\mu\nu}=0$ as in flat spacetime. A solution which generalizes the constancy of $\theta^{\mu\nu}$ in flat spacetime to curved spacetime is 
\begin{align}
 D_\lambda \theta^{\mu\nu} = 0. \nn
\end{align}
This implies $D_\lambda \xi^\mu \theta^{\lambda\nu} - D_\lambda \xi^\nu \theta^{\lambda\mu} = 0$ and the solution is $ D_\mu \xi^\mu = 0$. Hence there is a residual symmetry similar to what happens in flat spacetime. We are left with the symplectic subgroup of the volume preserving diffeomorphisms which also preserves a covariantly constant $\theta^{\mu\nu}$. This is our extension of noncommutativity to curved spacetime. We assume that $\theta^{\mu\nu}$ is a covariantly constant tensor. We can then define all geometric objects like the Christoffel symbol, Riemann tensor and so on in the usual way. 

We can now couple $\theta^{\mu\nu}$ to the Riemann tensor to build a noncommutative contribution to the usual Einstein-Hilbert action. If we go up to second order in $\theta^{\mu\nu}$ we find that the only independent combination is 
\begin{align} 
S_{NC} = \frac{1}{16\pi} \int d^4 x \,\, \sqrt{-g} \,\, \theta^{\mu\nu}
{\theta^{\alpha}}_\beta {R_{\mu\nu\alpha}}^\beta.\nn
\end{align}
The local symmetry is now reduced to volume preserving transformations. Gravity theories invariant under volume preserving transformations are well known and are called unimodular gravity \cite{Alvarez:2005iy}. Its main property is that the determinant of the metric is constant
and can be chosen to be one, the raison d'être of its name. Also the cosmological constant appears as an integration constant. Another interesting property is that the Einstein-Hilbert action has a finite polynomial form in the metric. 

We can derive the linearized equations of motion in the noncommutative case
\begin{align}
&\frac{1}{G} \left[ \Box
h_{\mu\nu} + \partial_\mu \partial_\nu
  {h_\rho}^\rho - \partial^\rho \partial_\mu h_{\nu\rho} -
  \partial^\rho \partial_\nu h_{\mu\rho} -  \eta_{\mu\nu} \left( \Box
    {h_\rho}^\rho - \partial^\rho \partial^\sigma h_{\rho\sigma}
    \right) \right] \nn \\
+ \theta^{\alpha\beta} \theta_{\mu\gamma} & \left(
            \partial_\nu \partial_\alpha {h_\beta}^\gamma - 
            \partial^\gamma \partial_\alpha h_{\beta\nu} \right) +
          \theta^{\alpha\beta} \theta_{\nu\gamma} \left( 
            \partial_\mu \partial_\alpha {h_\beta}^\gamma -
            \partial^\gamma \partial_\alpha h_{\beta\mu} \right) 
+ 2 \eta_{\mu\nu} \theta^{\alpha\beta} {\theta^\rho}_\sigma
\partial_\rho \partial_\alpha {h_\beta}^\sigma = 0, \nonumber
\end{align}
to find out that 
\begin{align}
g_{00} =& 1 + \left( 1 + 3 G \vec{\theta}^2  - 
G (\vec{n} \cdot \vec{\theta})^2 \right) h, \nn\\
g_{0i} =& 0, \nn\\
g_{ij} =& - \delta^{ij} + \left[ n^i n^j + G \delta^{ij}
  \left( \vec{\theta}^2 - (\vec{n} \cdot \vec{\theta})^2 \right) +
  G \left(\theta^i n^j + \theta^j n^i \right) \vec{n} \cdot
  \vec{\theta} \right] h, \nn
\end{align}
where $ h = -2GM/r$. We can find the correction to the Newtonian potential felt by a test particle 
\begin{align}
\frac{d^2 x^i}{dt^2} = - \frac{1}{2} \partial_i \left[ h + 
G \left( 3 \vec{\theta}^2 - (\vec{n} \cdot \vec{\theta} )^2 \right) h
\right]. \nn
\end{align}
The Newtonian potential has a contribution proportional to $\vec{\theta}^2$ which can be regarded as giving rise to an  effective Newton constant $G(1 + \frac{3}{2} G \vec{\theta}^2)$. The angular dependent piece $(\vec{n} \cdot \vec{\theta} )^2$ also contributes to the potential but the potential still goes as $1/r$. The force on a test particle is given by 
\begin{align}
m \frac{d^2 x^i}{dt^2} = m \left[ n^i + 3 G \left(
    \vec{\theta}^2 - (\vec{n} \cdot \vec{\theta})^2 \right) n^i + 
2 G (\vec{n} \cdot \vec{\theta}) \theta^i \right] \frac{h}{2r},\nn
\end{align}
so it is not purely radial. The radial component is modified by $3G(\vec{\theta}^2 - (\vec{n} \cdot \vec{\theta})^2$. The term proportional to $\vec{\theta}$ produces, in general, a force off the plane of the orbit. It is also periodic for closed orbits. If $\vec{\theta}$ is perpendicular to the plane of the orbit then no periodic effect due to noncommutativity is seen. 

Noncommutativity can be incorporated into gravity in alternative ways \cite{Nicolini:2008aj} and also in cosmological models and there is a huge literature on the subject \cite{GarciaCompean:2001wy,Barbosa:2004kp,Bastos:2008sg}. 

\section{Conclusions}

Noncommutative theories is still a broad area of research in physics and mathematics with applications ranging from condensed matter physics to particle physics and cosmology. We presented a short review on some topics of noncommutativity in field theory and gravity. Noncommutative field theories can be easily written by replacing the ordinary product among fields by the Moyal product. Other more complicated recipes also do exist. The main consequence of the Moyal product is the appearance of the UV/IR divergence which usually spoils renormalizability. Many proposal to overcome this problem were presented and several are still under study. Noncommutative gravity is a very broad area of research. Many proposals for it are available and we discussed just one of them. Also, gravity as an emergent phenomena can be realized in the noncommutative context, at the field theory level and also at the matrix model level. Many other application of noncommutativity were not mentioned here and are still an active area of research. 

\ack
 
I would like to thank the organizers of the XIV Mexican School on Particles and Fields for the kind invitation to deliver this talk and specially to David Vergara for hospitality. 
The work of V.O.R. is supported by CNPq grant No. 304495/2007-7 and FAPESP grant No. 2008/05343-5.

\section*{References}

\bibliography{noncommutative}

\end{document}